\documentclass[conference]{IEEEtran}
\IEEEoverridecommandlockouts
\usepackage{cite}
\usepackage{amsmath,amssymb,amsfonts}
\usepackage{algorithmic}
\usepackage{graphicx}
\usepackage{textcomp}
\usepackage{csquotes}
\usepackage{breqn}
\usepackage{xcolor}
\def\BibTeX{{\rm B\kern-.05em{\sc i\kern-.025em b}\kern-.08em
    T\kern-.1667em\lower.7ex\hbox{E}\kern-.125emX}}
\begin{document}

\title{Multi-reference Cosine: A New Approach to Text Similarity Measurement in Large Collections}

\author{\IEEEauthorblockN{Hamid Mohammadi}
\IEEEauthorblockA{\textit{Computer Engineering Department} \\
\textit{K.N.Toosi University of Technology}\\
Tehran,  Iran \\
mohammadi2823@email.kntu.ac.ir}
\and
\IEEEauthorblockN{Amin Nikoukaran}
\IEEEauthorblockA{\textit{Computer Engineering Department} \\
\textit{University Of Tehran}\\
Tehran,  Iran \\
aminnikookaran@ut.ac.ir}
}

\maketitle

\begin{abstract}
The importance of an efficient and scalable document similarity detection system is undeniable nowadays. Search engines need batch text similarity measures to detect duplicated and near-duplicated web pages in their indexes in order to prevent indexing a web page multiple times. Furthermore, in the scoring phase, search engines need similarity measures to detect duplicated contents on web pages so as to increase the quality of their results. \\
In this paper, a new approach to batch text similarity detection is proposed by combining some ideas from dimensionality reduction techniques and information gain theory. The new approach is focused on search engines need to detect duplicated and near-duplicated web pages. The new approach is evaluated on the NEWS20 dataset and the results show that the new approach is faster than the cosine text similarity algorithm in terms of speed and performance. On top of that, It is faster and more accurate than the other rival method, Simhash similarity algorithm.
\end{abstract}


\section{Introduction}
\label{sec:1}
Nowadays, one of the basic and critical abilities of a search engine is to detect similarity between documents or more precisely between web pages on the internet. Search engines use text similarity algorithms for tasks such as indexing or web page ranking \cite{conrad2003online, fetterly2004spam}. There are two main causes of the similarity between web pages. The first cause is that there are always many links to a single web page. These links could be created by search engine optimization companies. The second problem is that there are always some web pages, which have nearly similar contents compared to other more popular web pages. In another word, there are lots of duplicated contents on the world wide web \cite{henzinger2007search}. By detecting this similarity, search engines can have a good understanding of the freshness and originality of web pages. One of the greatest difficulties of internet document similarity detection is the huge number of documents available on the web and it is rapid growth. The problem of old document similarity detection algorithms was that they have to compare all documents available to them with each other so as to detect duplicated or near-duplicated documents. In order to compare each new web page with all the other web pages, the algorithm needs a really massive amount of storage space and processing power. In order to solve this problem, Brin et al. \cite{brin1995copy} introduced a novel approach to overcome the difficulties of this process. Subsequently, signature-based document similarity detection algorithms, like Simhash, used widely by search engines like Google to detect duplicated or near-duplicated web pages \cite{manku2007detecting}. The advantage of hashing algorithms over other text similarity detection approaches, like cosine text similarity or Jaccard text similarity, is that it can compare documents using only their small sized hashes instead of the whole documents \cite{stein2007principles, potthast2008new}. Hash-based duplicated and near-duplicated document detection methods create a hash database of documents available on the internet or any interested dataset and then detect similar hashes based on their hamming distance \cite{stein2007principles, sood2012probabilistic}. Hashing algorithms have their own disadvantages. By reducing document's size, for instance, to a 64 bits hash, they lose a lot of essential information about a document, and therefore they lose their similarity detection accuracy compared to algorithms such as cosine or Jaccard similarity \cite{sood2012probabilistic}. The importance of this paper is the proposal of a new algorithm that has both cosine text similarity accuracy and hash-based algorithms ability to generate a signature for texts and compare them using that signature. This paper is organized as follows: Section \ref{sec:2} discusses related works. Section \ref{sec:3} describes the cosine and Simhash similarity measures. Next, section \ref{sec:4} proposes the new approach. The test results are shown in section \ref{sec:5} and finally, section \ref{sec:6} presents the conclusions.

\section{Related works}
\label{sec:2}
There are three main approaches to duplicated and near-duplicated web page detection problem: Semantic, URL based and syntactic \cite{alsulami2012near}. Semantic approaches are based on the meaning of the text on the web page. One of the semantic approaches to web page similarity is fuzzy semantic-based string similarity \cite{alzahrani2010fuzzy}. In this approach, after pre-processing the texts on web pages and collecting a list of possible near-duplicated web pages using shingling and Jaccard algorithms, the web pages are compared using sentence-wise algorithms and a fuzzy similarity degree will be given to the pages. Another approach to duplicated and near-duplicated webpage detection is URL based approaches. These approaches are focused on detecting web pages that are possibly duplicated or near-duplicated based on their URL relations. One novel algorithm in this field is the Dust Buster \cite{bar2009not}. To achieve this algorithm, they analyzed the crawl and web server logs to discover the rules of detecting web pages with similar content from their URLs. The other approach to duplicated and near-duplicated web-page detection is the syntactic approach. One of the first approaches of this type to detect duplicated and near-duplicated documents was shingling \cite{broder1997syntactic}. In this approach, overlapping fragments called shingles are created and the similarity degree is determined by counting the equal shingles. The other algorithm in the field of syntactic text similarity measurement is pair-wise similarity computation \cite{alsulami2012near}, which is an algorithm focused on finding pairs of similar documents in a large text document collection, using some similarity measures. SpotSigs \cite{theobald2008spotsigs} is another syntactic approach to duplicated and near-duplicated web page detection. SpotSigs is an approach to find more important shingles when comparing web pages, so the comparison result will be more accurate. Machine learning plays a great role in text similarity measurements mainly in semantic and syntactic comparison. In the semantic domain, the SemSim \cite{kashyap2016robust} is a good example of machine learning application in text similarity. SemSim uses semantic features and Support Vector Machines to measure the semantic similarity between two texts.

\section{Similarity measures}
\label{sec:3}
With the advent of the World Wide Web in 1990, the need for information retrieval systems grew up rapidly. One of the important tools of any information retrieval system is document similarity detection algorithms. There are two popular algorithms that are used for document similarity detection applications, referred to as cosine and Simhash algorithms. Cosine text similarity \cite{lukashenko2007computer, singhal2001modern} is based on measuring the cosine of the angle between two vector space models \cite{liang2010improved} created from two text documents. The Simhash algorithm is based on generating a signature for each document and comparing documents using their signatures. \\
Cosine text similarity algorithm calculates the similarity between two text documents by counting all N-grams (e.g. 3-grams) available in the documents and stores them in an M-dimensional vector, where M is the number of unique N-grams found in the passage. Cosine text similarity algorithm uses this vector to compare documents. The cosine text similarity algorithm compares the vectors by calculating the angle between two vectors using Eq. \ref{eq:1}.

\begin{dmath}
\label{eq:1}
cosine\ text\ similarity = \frac{\overrightarrow{A}.\overrightarrow{B}}{|A||B|}
\end{dmath}

Where A and B are vector space models of two documents. The value of cosine text similarity is a decimal number between 0 and 1, which 0 indicates not similar at all and 1 indicates exact similarity and values in-between indicate the similarity degree. Cosine similarity algorithm has some advantages and disadvantages. One advantage of the cosine similarity algorithm is that the algorithm is independent of the language of the document. The algorithm can use 3-grams to compare text instead of the meaning of the words. Another advantage of cosine algorithms is that its result is not related to the order of the words in a document. These features give a cosine similarity measure the ability to detect plagiarism \cite{lukashenko2007computer}. Generally, cosine text similarity measure has better recall and precision than most of other text similarity measures like Pearson Correlation Coefficient (PCC), Jaccard or Mean Square Difference (MSD) according to this research \cite{liu2014new}. \\
Simhash is a text hashing algorithm based on cosine text similarity algorithm \cite{shrivastava2014defense}. This algorithm has good precision and recall and has been used as a near-duplication detection method in real-world problems in search engines like Google. In Simhash approach, the algorithm tokenizes the input text and extracts words out of it. Next, it hashes the words using common hashing algorithms like MD5 or SHA1 into N bits hashes. Furthermore, it puts bit representation of all created hashes in a column and for each column of bits, it changes each 0 value to -1. Then, it calculates the sum of all values in each bit column and if the summation of each column was bigger than 0 value, then it assumes 1 for this column summation result otherwise it sets the value 0.  After repeating the recent operation for each column, the final result of this operations is a hash value with N bits length. In order to compare different hash values, Simhash uses hamming distance. The number of differences between two binary strings when comparing corresponding bits is the Hamming distance between those two strings. The Hamming distance between two documents represents the degree of difference between them. \\
In addition, simhash has some interesting applications. For example, it can be used in a privacy-preserving recommendation system \cite{xu2017privacy} or it is also useful in similarity detection of short texts \cite{zhang2017research}.

\section{Proposed algorithm}
\label{sec:4}
\begin{figure}
	\centering
    	\includegraphics[width=0.45\textwidth]{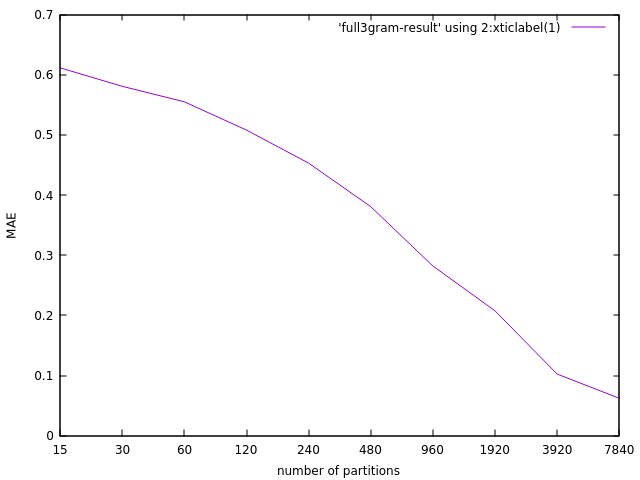}
	\caption{Mean absolute error of the proposed approach}
    	\label{fig:1}
\end{figure}

\begin{figure}
	\centering
    	\includegraphics[width=0.45\textwidth]{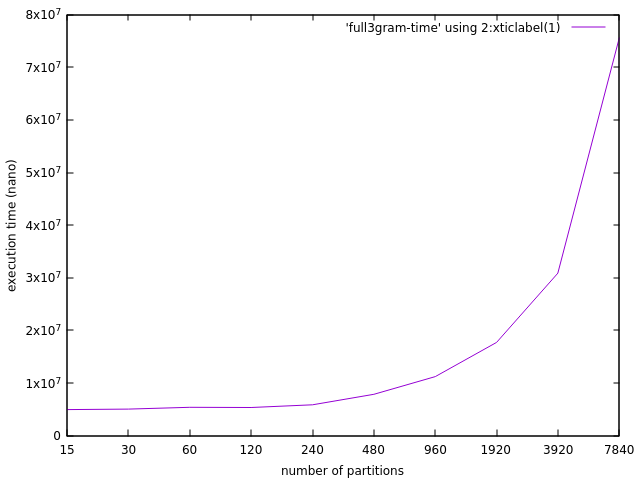}
	\caption{Execution time of the proposed approach}
    	\label{fig:2}
\end{figure}

We combined the idea of hash-based text similarity algorithms and cosine text similarity to create a signature based text similarity algorithm. In order to create a signature for a document, the proposed algorithm compares each document with a reference string using cosine text similarity algorithm. The main part of the algorithm is how we create the reference and generate a signature using that reference. Different references texts can be created and each reference text has its own advantages and disadvantages. There is a trade-off between the performance and the accuracy of the algorithm using different references. This trade-off gives the proposed algorithm a good flexibility to be applied to different applications. We can tune reference texts base on the need to achieve more accuracy with lower performance or better performance with less accuracy. \\
The simplest form of reference could be a list of all possible 3-grams. In the new approach, an N-gram is an N-alphabet sequence string. We concatenate these 3-grams and generate a string as a reference. By using this simple reference string, the algorithm generates nearly similar signatures for all documents. Since this reference has all the 3-grams possible, all documents are similar to this reference. Each document has a different degree of similarity to each part of the reference text. Therefore, by splitting the reference text into a number of substrings with the equal length it is possible to improve the efficiency of the reference text. By changing the number of reference partitions, it is possible to target a smaller number of 3-grams, so the overall comparison will be more precise. Figure \ref{fig:1} and Figure \ref{fig:2} show the Mean Absolute Error (MAE) and execution time of the approach, respectively.

\begin{figure}
	\centering
    	\includegraphics[width=0.45\textwidth]{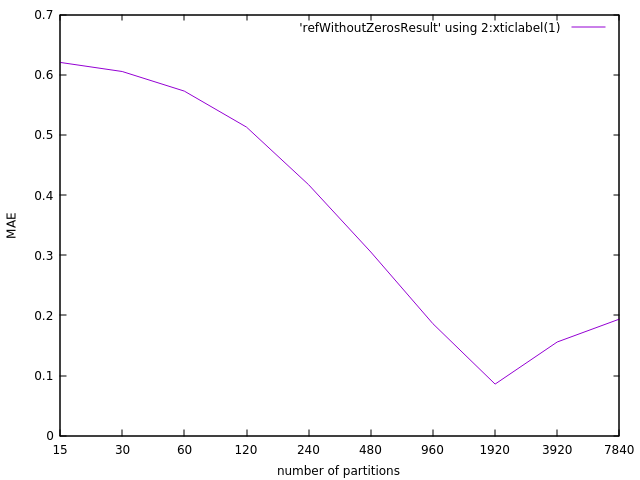}
	\caption{Mean absolute error of multi-reference cosine with zero frequency 3-grams removed reference}
    	\label{fig:3}
\end{figure}

\begin{figure}
	\centering
    	\includegraphics[width=0.45\textwidth]{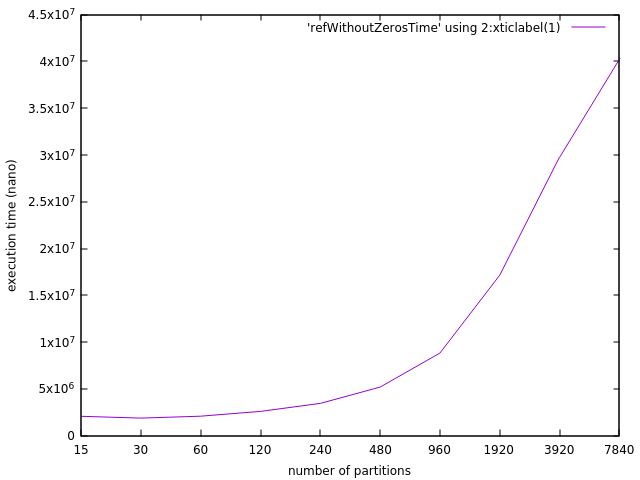}
	\caption{Mean absolute error of multi-reference cosine with zero frequency 3-grams removed reference}
    	\label{fig:4}
\end{figure}

According to Figure \ref{fig:1}, by increasing the number of reference text partitions, the mean absolute error of the proposed algorithm from cosine text similarity will move toward zero. This result indicates that there is a close relationship between the multi-reference cosine and the cosine text similarity algorithm. Figure \ref{fig:2} shows that the proposed algorithm needs an array of float values with the length of 7840 and $7.5 \times 10_{7}$ nanoseconds of time in order to achieve a mean absolute error of 0.062. Aside from showing the correlation between the proposed algorithm and the cosine text similarity algorithm, these results are unacceptable but it is possible to further improve the results. \\
3-grams of English language is not uniformly distributed according to their frequency. For example, there are about 9000 3-grams in Open National American Corpus that had almost zero frequency and their effect on text similarity detection is negligible. Furthermore, the importance of some 3-grams is less than the others. In addition to that, the 3-grams \enquote{the} and \enquote{and} are so frequent that almost all document are similar if they are compared according to the existence of these 3-grams. On the other hand, 3-grams like \enquote{aaa} or \enquote{zzz} are so rare that almost all documents are similar to each other in terms of not having them. It is possible to reduce reference text size and execution time by choosing the 3-grams, which have a higher importance in similarity detection than the others. To create such reference, Open American National Corpus is analyzed \cite{ide2008american} so as to calculate 3-grams frequencies. Using this information, the new approach achieves better signatures that have higher precision and lesser mean absolute error from the cosine similarity algorithm. The new results have acceptable time and signature size. \\
By generating a list of English 3-grams that are sorted by their frequencies, 3-grams that have zero frequency are removed. It is observed that the new approach obtained slightly more speed and higher precision with fewer partitions (Figure \ref{fig:3} and \ref{fig:4}).

\begin{figure}
	\centering
    	\includegraphics[width=0.45\textwidth]{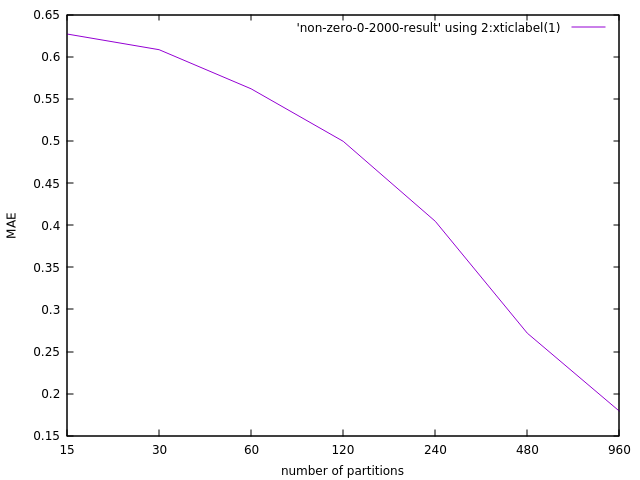}
	\caption{Mean absolute error of multi-reference cosine – (0 to 2000)}
    	\label{fig:5}
\end{figure}

\begin{figure}
	\centering
    	\includegraphics[width=0.45\textwidth]{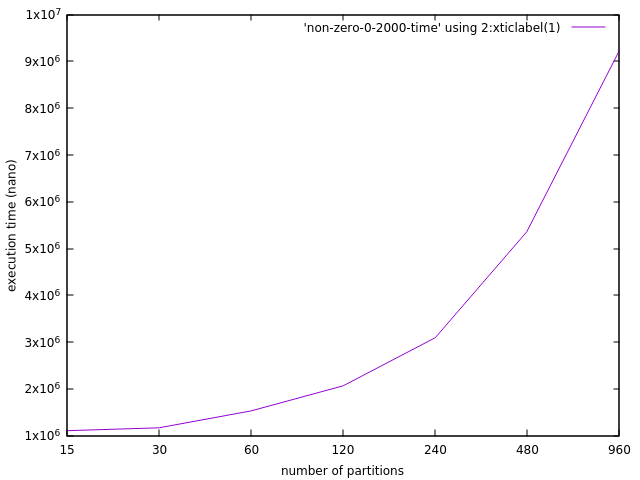}
	\caption{Execution time of multi-reference cosine – (0 to 2000)}
    	\label{fig:6}
\end{figure}

\begin{figure}
	\centering
    	\includegraphics[width=0.45\textwidth]{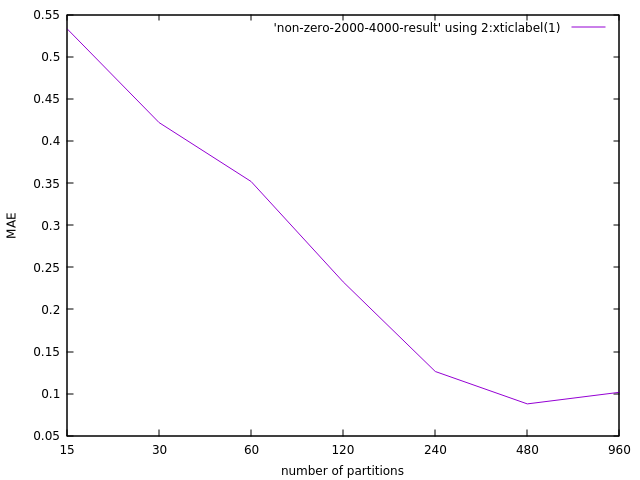}
	\caption{Mean absolute error of multi-reference cosine –(2000 to 4000)}
    	\label{fig:7}
\end{figure}

\begin{figure}
	\centering
    	\includegraphics[width=0.45\textwidth]{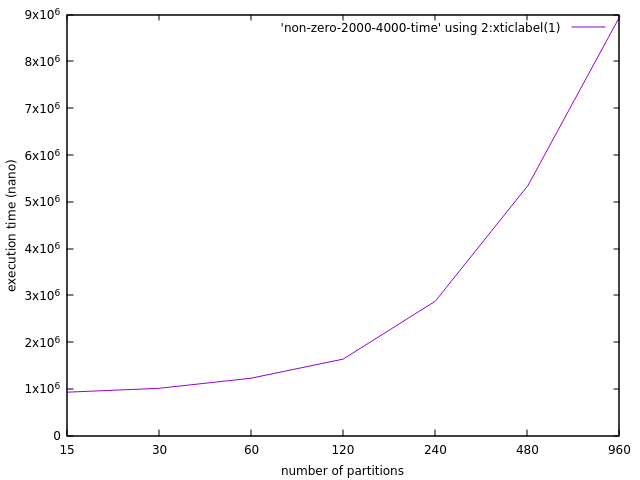}
	\caption{Execution time of multi-reference cosine – (2000 to 4000)}
    	\label{fig:8}
\end{figure}

\begin{figure}
	\centering
    	\includegraphics[width=0.45\textwidth]{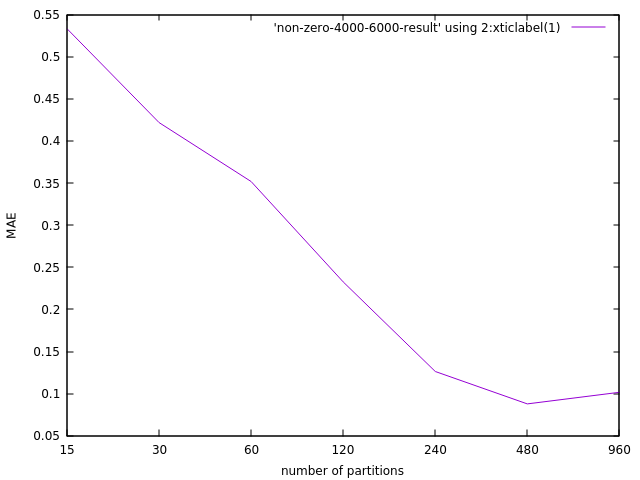}
	\caption{Mean absolute error of multi-reference cosine – (4000 to 6000)}
    	\label{fig:9}
\end{figure}

\begin{figure}
	\centering
    	\includegraphics[width=0.45\textwidth]{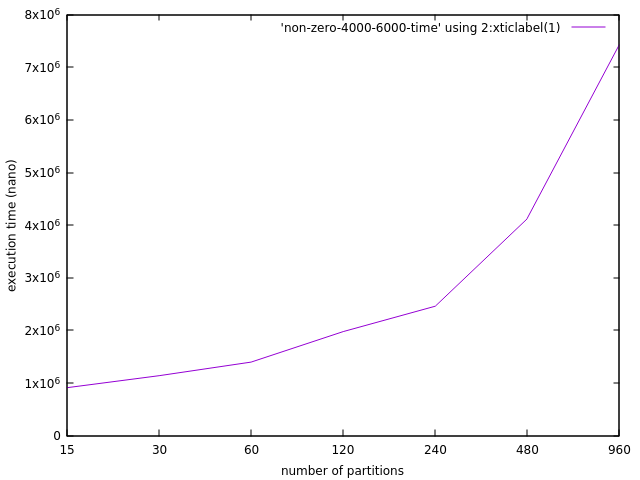}
	\caption{Execution time of multi-reference cosine – (4000 to 6000)}
    	\label{fig:10}
\end{figure}

\begin{figure}
	\centering
    	\includegraphics[width=0.45\textwidth]{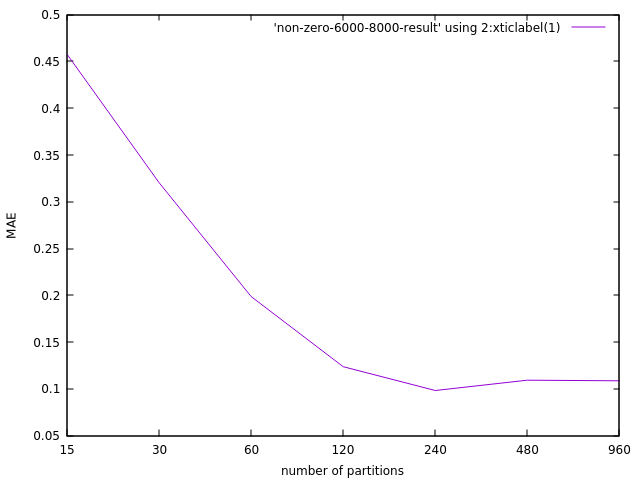}
	\caption{Mean absolute error of multi-reference cosine – (6000 to 8000)}
    	\label{fig:11}
\end{figure}

\begin{figure}
	\centering
    	\includegraphics[width=0.45\textwidth]{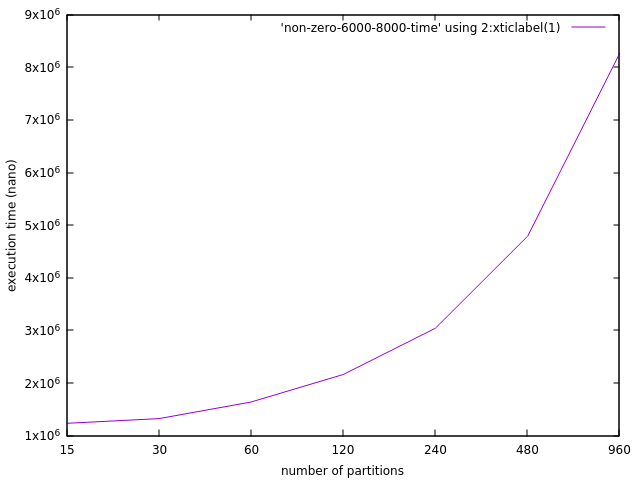}
	\caption{Execution time of multi-reference cosine – (6000 to 8000)}
    	\label{fig:12}
\end{figure}

As shown in Figure \ref{fig:3} and Figure \ref{fig:4}, there is a slight improvement both in speed and mean absolute error. The approach reached the mean absolute error of 0.0858 by splitting the reference into 1920 parts. The best mean absolute error in the last experiment (Figure \ref{fig:1}) was 0.062. Considering the time taken for the new approach, by removing the zero frequency 3-grams from references, the performance is improved. For example, the time taken is $4 \times 10_{8}$ nanoseconds for a 7840 parted reference but last experiment time consummation (Figure \ref{fig:2}) for 7840 parted reference was $7.5 \times 10_{7}$ nanoseconds. By selecting the more important 3-grams, which have more information about how similar or different two documents are, the algorithm can achieve better mean absolute error in less time, with smaller signatures. \\
The length of the 3-grams list, not taking into account zero frequency 3-grams, is about 8000. This list is sorted descending regarding the frequency of 3-grams. Four different reference texts are created, each containing 2000 3-grams, all from a specific frequency region of the 3-grams list. For example, the region (0 – 2000) and (6000 – 8000) respectively have most and least frequent 3-grams. The MAE and execution time of the new algorithm using the above references are shown in Figures \ref{fig:5} to \ref{fig:12}.

\begin{figure}
	\centering
    	\includegraphics[width=0.45\textwidth]{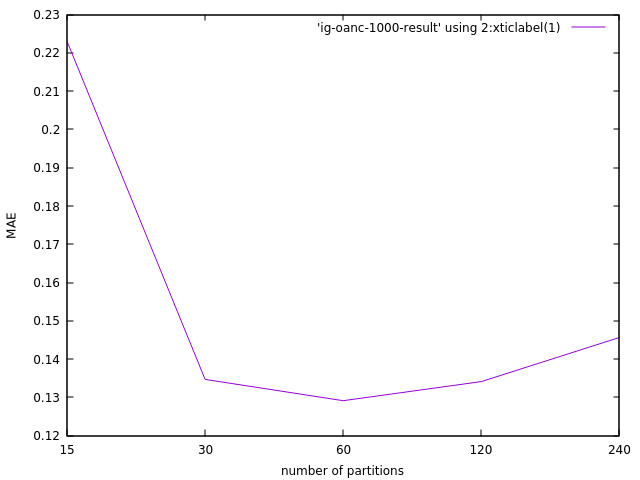}
	\caption{Mean absolute error of multi-reference cosine – 1000 3-grams}
    	\label{fig:13}
\end{figure}

\begin{figure}
	\centering
    	\includegraphics[width=0.45\textwidth]{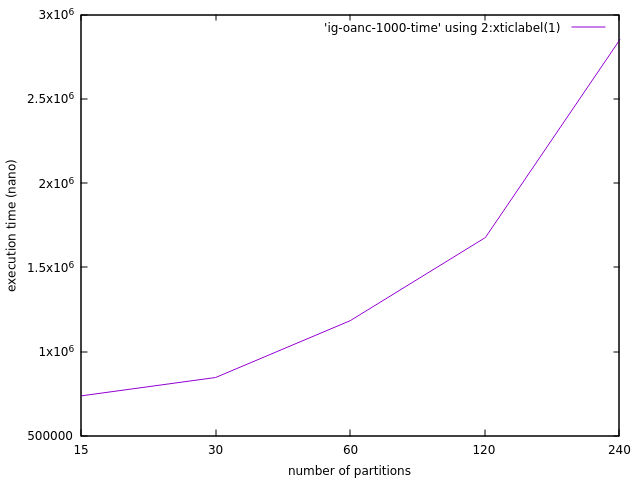}
	\caption{Execution time of multi-reference cosine – 1000 3-grams}
    	\label{fig:14}
\end{figure}

\begin{figure}
	\centering
    	\includegraphics[width=0.45\textwidth]{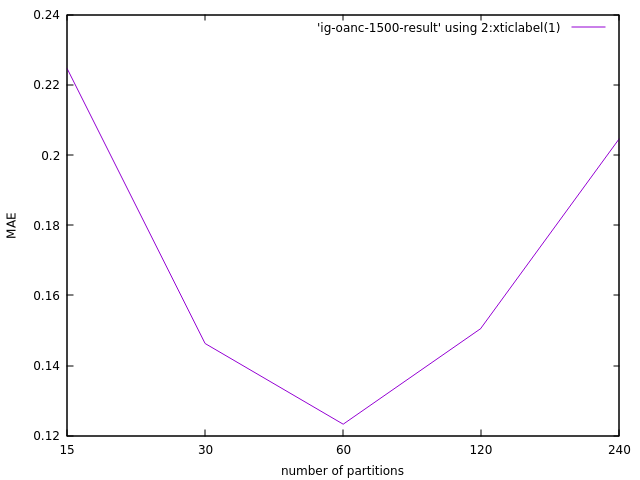}
	\caption{Mean absolute error of multi-reference cosine – 1500 3-grams}
    	\label{fig:15}
\end{figure}

\begin{figure}
	\centering
    	\includegraphics[width=0.45\textwidth]{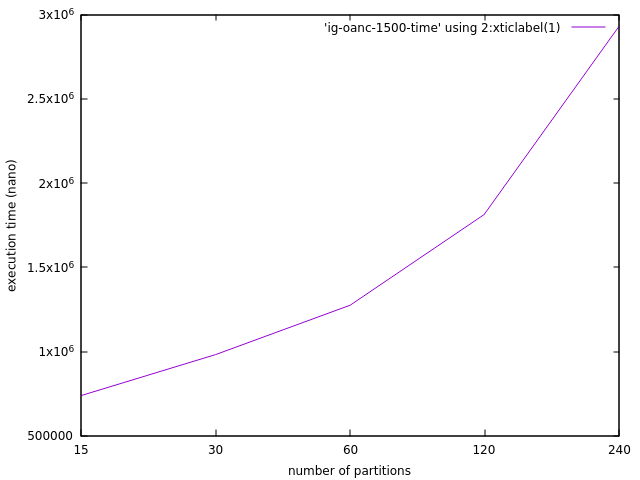}
	\caption{Execution time of multi-reference cosine – 1500 3-grams}
    	\label{fig:16}
\end{figure}

\begin{figure}
	\centering
    	\includegraphics[width=0.45\textwidth]{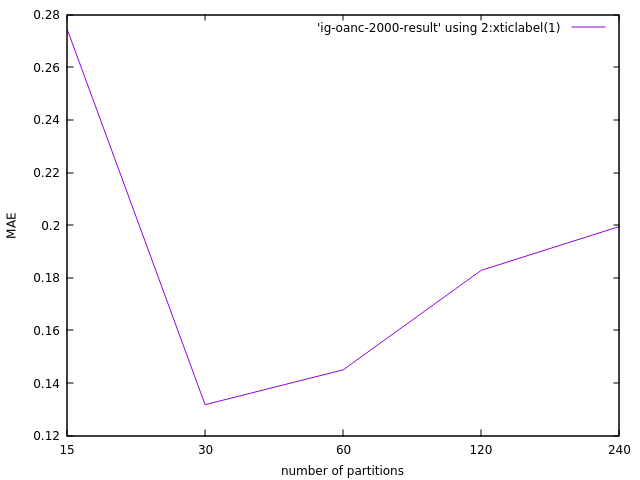}
	\caption{Mean absolute error of multi-reference cosine – 2000 3-grams}
    	\label{fig:17}
\end{figure}

\begin{figure}
	\centering
    	\includegraphics[width=0.45\textwidth]{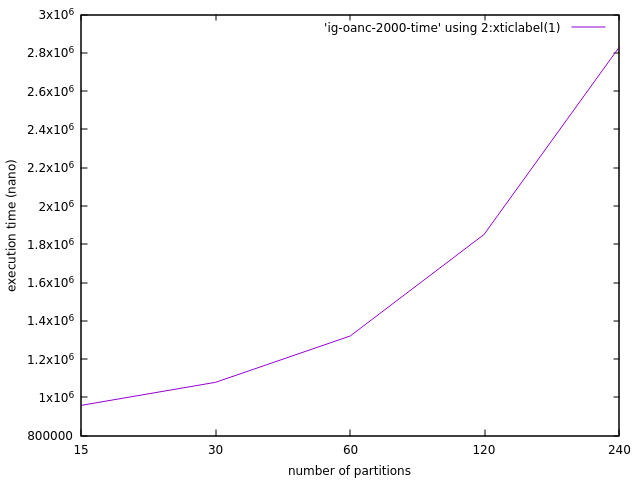}
	\caption{Execution time of multi-reference cosine – 2000 3-grams}
    	\label{fig:18}
\end{figure}

The new approach achieves less mean absolute error and less execution time by using 3-grams with less frequency. Using this reference text the new approach reaches good precision with a smaller signature size. For example, by choosing 2000 of least frequent 3-grams and splitting it into 120 partitions, the algorithm reaches the value 0.1236 for mean absolute error (Figure \ref{fig:11}) with approximately $2.16 \times 10_{6}$ nanoseconds of execution time (Figure \ref{fig:12}). By splitting the reference into more than 240 partitions, the algorithm precision decreases (Figure \ref{fig:11}). We concluded that the algorithm cannot achieve more precision just by increasing the number of reference partitions.
From a list of 3-grams, which were ordered by 3-grams frequency and zero frequency 3-grams are removed, 3-grams with the frequency of one are chosen. These 3-grams have the most significant impact on similarity or dissimilarity of two documents because of their uniqueness and higher information gain. Next, those 3-grams, which have low frequency and only existed in a single document and not in any other documents in the News20 dataset, are selected. Therefore, the algorithm can distinguish documents using these unique 3-grams because of their high information gain. By extracting a list of 1000, 1500 or 2000 of recently described 3-grams from the News20 dataset, it is possible to generate more effective references. In this experiment, recently described reference text is used. The precision and execution time of the algorithm using these references are shown in Figure \ref{fig:13} to Figure \ref{fig:18}.

By choosing 1500 important 3-gram with most information gain and then splitting them into 60 partitions, the algorithm obtains a better precision with less execution time comparing to the previous results. The algorithm can reach a mean absolute error of 0.1232 (Figure \ref{fig:15}) from cosine text similarity in a execution time less than $1.27 * 10^6$ nanoseconds (\ref{fig:16}). Additionally, it achieves a mean absolute error of 0.1317 (Figure \ref{fig:17}) with half of the number of reference parts and reduces 16 percent of execution time, which is very significant for big practical applications like duplicated and near-duplicated web page detection on the web. Also, by using the reference with 30 partitions, the algorithm loses a little bit of precision, but it can get better performance and it is possible to reduce the signature size by 50 percent.

\section{Test results}
\label{sec:5}
3-grams are used in order to create the reference texts for these tests because they provide better accuracy than unigrams and bigrams with N \textless 3 and better performance than N-grams with N \textgreater 3 due to fewer number of permutations possible for 3 character long strings (3-grams) made out of the language characters, therefore smaller N-grams space size \cite{kondrak2005n}. The error of the proposed approach can be computed using the following equations Eq. \ref{eq:2} and Eq. \ref{eq:3}:

\begin{dmath}
MAE = \frac{1}{T_{D}}\sum_{i,j \epsilon D}^N {(multi \ reference\ cosine(D_{i}, D_{j})} - cosine(D_{i}, D_{j}))
\label{eq:2}
\end{dmath}

\begin{dmath}
Variance = \frac{1}{T_{D}}\sum_{i,j \epsilon D}^N ( {multi\ reference\ cosine(D_{i}, D_{j})} - mean\ distance) ^ {2}
\label{eq:3}
\end{dmath}

Where D is the dataset, $D_{i}$, $D_{j}$ are $i^th$ and $j^th$ document in the dataset and $T_{D}$ is the total number of documents in the dataset. \\
To calculate the mean absolute error (MAE) and variance for the Simhash algorithm, Simhash results are used instead of multi-reference cosine results in the Eq. \ref{eq:2} and Eq. \ref{eq:3}. Variance shows how a document similarity algorithm oscillates around the ground truth values. All results are computed using an Intel Core i7-4510U CPU computer with 8 gigabytes of RAM, running java on OpenJDK JVM, version $1.8.0\_91$. \\
NEWS20 \cite{ide2008american} dataset is chosen for this experiment, which was collected originally by Ken Lang. Since the NEWS20 dataset has about 18828 different documents, this dataset is a reasonable dataset for document similarity tasks. First, each document in this dataset is compared to other documents using cosine text similarity algorithm, which is the ground truth. Then the multi-reference cosine results are compared to the cosine text similarity results. The algorithm uses a 1000 3-grams reference generated using Information gain theory over Open American National Corpus references with 30 and 60 partitions. Table \ref{tab:1} shows the results of this experiment:

\begin{table}
\caption{comparison between different text similarity algorithms}
\label{tab:1}       % Give a unique label
\begin{tabular}{p{0.45\linewidth}p{0.12\linewidth}p{0.1\linewidth}p{0.15\linewidth}}
\hline\noalign{\smallskip}
Algorithm & Mean Absolute Error & Variance & Average execution time ($\mu$μs) \\
\noalign{\smallskip}\hline\noalign{\smallskip}
Simhash & 0.18854 & 0.04173 & 327 \\
Proposed algorithm (1000 3-grams - 30 partitions) & 0.13471 & 0.02633 & 711 \\
Proposed algorithm (1000 3-grams - 60 partitions) & 0.12914 & 0.02107 & 941 \\
Proposed algorithm (1500 3-grams - 60 partitions) & 0.12328 & 0.01947 & 1027 \\
\noalign{\smallskip}\hline
\end{tabular}
\end{table}

As shown in Table \ref{tab:1}, the multi-reference cosine achieves less Mean absolute Error than Simhash similarity algorithm. The algorithm also has less variance. Less variance shows that the proposed approach results are distributed within a smaller distance from cosine similarity results, therefore it has more reliable results.

\section{CONCLUSIONS AND FUTURE WORKS}
\label{sec:6}
In this paper, a new similarity measure is proposed to detect similarity between different documents which is accurate, fast and can produce small-sized signatures (about 240 to 480 bytes). The new approach has precision close to the cosine text similarity algorithm alongside with the signature generation and signature base comparison ability. This algorithm has various application of document similarity, plagiarism detection and clustering. The new approach is tested on the NEWS20 dataset and compared to Simhash and cosine text similarity algorithms. With the same time consumption of Simhash algorithm, the new approach has reached better results in terms of accuracy and variance. Despite its good performance and precision, as a future work, it is possible to improve the algorithm by using genetic algorithm, to generate more effective references. It is also possible to use weighted vector space models in order to achieve more accuracy with smaller references.

\bibliographystyle{IEEEtran}
\bibliography{refs}

\end{document}